\documentclass[10pt,letterpaper,american]{article}
\usepackage[latin9]{inputenc}
\usepackage{color}
\usepackage{babel}
\usepackage{amsmath}
\usepackage{amssymb}
\usepackage{graphicx}
\usepackage[bookmarks=false,
 breaklinks=false,pdfborder={0 0 1},backref=false,colorlinks=true]
 {hyperref}
\hypersetup{
 citecolor=blue,urlcolor=blue}

\makeatletter

\pdfpageheight\paperheight
\pdfpagewidth\paperwidth

\usepackage{osameet3}%% use version 3 for proper copyright statement
\usepackage{capt-of}
\usepackage{booktabs}

\usepackage{cite}
\usepackage[printonlyused]{acronym}
\acrodef{DM}{distribution matcher}
\acrodef{Hi-DM}{hierarchical DM}
\acrodef{PS}{probabilistic shaping}
\acrodef{PAS}{probabilistic amplitude shaping}
\acrodef{AWGN}{additive white Gaussian noise}
\acrodef{QAM}{quadrature amplitude modulation}
\acrodef{FEC}{forward error correction}
\acrodef{PAM}{pulse amplitude modulation}
\acrodef{LUT}{look up table}
\acrodef{MB}{Maxwell-Boltzmann}
\acrodef{ESS}{enumerative sphere shaping}
\acrodef{CCDM}{constant composition distribution matching}
\acrodef{IR}{information rate}
\acrodef{SNR}{signal to noise ratio}
\acrodef{PPM}{pulse position modulation}
\acrodef{invDM}{inverse DM}
\acrodef{TX}{transmitter}
\acrodef{RX}{receiver}
\acrodef{BER}{bit error rate}
\acrodef{SER}{symbol error rate}

\makeatother

\begin{document}
\title{Nonlinearity Mitigation for Coherent Ground-to-Satellite Optical Links}
\author{S. Civelli\textsuperscript{1,2,{*}}, M. Secondini\textsuperscript{2,3},
L. Pot\`i\textsuperscript{3,4}}
\address{\textsuperscript{1}CNR-IEIIT, Via Caruso 16, 56122, Pisa, Italy,
\emph{stella.civelli@cnr.it}\\
\textsuperscript{2} TeCIP Institute, Scuola Superiore Sant'Anna,
Via G. Moruzzi 1, 56124, Pisa, Italy\\
\textsuperscript{3}PNTLab, CNIT, Via G. Moruzzi 1, 56124, Pisa, Italy\\
\textsuperscript{4}Universitas Mercatorum, Piazza Mattei 10, 00186,
Roma, Italy}

\maketitle
%% Uncomment the following line to override copyright year from the default current year.
\copyrightyear{2025}
\begin{abstract}
We propose digital signal processing techniques for nonlinearity mitigation
in high power optical amplifiers used in satellite communications.
The acceptable link loss increases by $6$dB with negligible complexity.\vspace*{-1ex}
\end{abstract}

\section{Introduction }

\vspace*{-1ex}

Recently, optical satellite communication has emerged as an effective
technology to complement---and potentially replace---conventional
radio-frequency (RF) satellite links \cite{kaushal2016optical,li2022survey,boddeda2023achievableJLT,boddeda2024current}.
Ground-to-satellite optical links, however, suffer from severe impairments
caused by time-varying atmospheric conditions and by the relative
motion between the satellite and the Earth (i.e., varying pointing
angle), both leading to significant attenuation and performance degradation.
To ensure sufficient received power at the satellite, thereby improving
throughput and maintaining resilience under adverse conditions, the
launch power should be as high as possible. Recently, several high-power
optical amplifiers (HPOAs) have been demonstrated, including in-band
core-pumped Er-doped fibers \cite{kotov2022high}, cladding-pumped
Er-doped fibers \cite{kotov2014yb}, and cladding-pumped ErYb co-doped
fibers \cite{matniyaz2020302}. All these approaches exhibit physical
and practical constraints or limitations, making custom designs necessary
depending on the specific application \cite{nicholson2025high}. Among
the main challenges, fiber nonlinearities invariably arise when a
high-power signal propagates through optical fibers---either within
the doped fiber inside the amplifier or along the output pigtail or
any additional patch cords. Its mitigation is essential to increase
the launch power, thereby enabling higher throughput or longer transmission
distances \cite{billault2023optical,ciaramella2024nonlinear}. On
one hand, the nonlinearity can be mitigated by carefully designing
the HPOA. For instance, erbium-doped fiber amplifiers (EDFAs) typically
require relatively longer doped fiber compared to erbium-ytterbium
codoped amplifiers (EYDFAs), making the latter a better options to
counteract nonlinear effects. However, very high power EYDFAs suffer
from higher-order mode power transfer leading to pointing stability
fluctuation as well as long term power degradation due to photodarkening
\cite{nicholson2025high}. In addition, the large-mode-area approach
lacks commercially available and reliable components, and maintaining
single-mode operation can be challenging. On the other hand, fiber
nonlinearity can be mitigated by properly designing the signal and/or
using digital signal processing (DSP). For example, the Kerr nonlinearity
occurring in a short single-mode fiber (SMF) pigtail following an
HPOA has been studied in \cite{ciaramella2024nonlinear}, where a
compensation method for direct detection systems was proposed. Furthermore,
coherent communications have become a reality for space communications,
enabling more advanced DSP and higher throughput \cite{li2022survey,guiomar2022coherent}.
In this paper, we propose mitigation and compensation techniques for
coherent optical satellite communication uplinks, introducing low-complexity
DSP methods that enable higher launch power and, consequently, longer
transmission distances or higher data rates.

\section{Nonlinearity Mitigation Techniques for High Power Optical Amplifier}

\vspace*{-1ex}

In this paper, we focus only on Kerr nonlinearity, expected to be
the major issue \cite{ciaramella2024nonlinear}. While the mitigation
of Kerr nonlinearity in optical fiber communication systems has been
extensively investigated, the dynamics in an HPOA are different. In
the former case, nonlinearity interacts with dispersion over a long
time duration (proportional to the symbol rate and fiber length),
giving rise to self- and cross-phase modulation effects, and some
four-wave mixing. The spectral broadening is minimal and the optical
power is relatively low (typically below 20\,dBm). In contrast, the
fiber within an HPOA is short ($<100$\,m), but the optical power
is much higher (up to $\approx50$dBm). Under these conditions, nonlinearity
interacts only weakly with dispersion, depends mostly on the instantaneous
power, and manifests mainly as spectral broadening. In this work,
we propose two DSP techniques, inherited from optical fiber communication
but specifically tailored to the HPOA scenario, to mitigate nonlinearity:
signal shaping and nonlinear phase rotation (NLPR).

Probabilistic amplitude shaping (PAS) implements constellation shaping
using a distribution matcher (DM), which maps information bits on
blocks of $N$ \emph{shaped} amplitudes, with a desired probability
distribution \cite{bocherer2015bandwidth}. For limited block lengths
$N$, the amplitudes within each block are strongly correlated. This
correlation induces a rate loss, which vanishes as $N$ increases.
At the same time, it reduces amplitude fluctuations within the block,
leading to a corresponding reduction of nonlinear effects. Sphere
shaping with 4D mapping (across quadratures and polarizations) often
achieves the best performance, with the optimal block length depending
on the channel memory induced by fiber dispersion. While in fiber
systems the optimal $N$ is generally large ($N>64$) \cite{civelli2022JLTBPS},
in the HPOA scenario the impact of fiber dispersion is negligible,
and the optimal block length is expected to be considerably shorter.
In this work, we propose using sphere shaping with 4D mapping and
a very short block length ($N=4$), which, as we will show, provides
significant shaping gain and can be readily implemented via a LUT
with negligible complexity.

Next, we propose to apply a nonlinear phase rotation (NLPR) to mitigate
nonlinearity. Neglecting the impact of dispersion, the effect of fiber
nonlinearity can be modeled as a simple phase rotation $\theta(t)$
proportional to the instantaneous signal power . This distortion
can be compensated digitally by applying the inverse phase rotation
$-\theta(t)$ at the TX, the RX, or distributed between both, with
very low computational complexity.  Performing the NLPR at the TX
would be most beneficial as (i) the transmitted signal is noiseless,
avoiding signal--noise interaction, and (ii) the satellite station
has very strict requirements in terms of complexity. However,\textcolor{red}{{}
}both the transmitter and receiver front ends are bandlimited. Due
to the significant spectral broadening introduced by nonlinearity,
portions of the signal spectrum may be filtered out, reducing the
overall effectiveness of the NLPR. To address this issue, it can be
beneficial to split the NLPR between the TX ($-\kappa\theta$) and
the RX ($(\kappa-1)\theta$), with $\kappa$ to be optimized. As we
will show, this approach mitigates spectral broadening and improves
the overall performance.

\begin{figure}
\centering

\includegraphics[width=1\columnwidth]{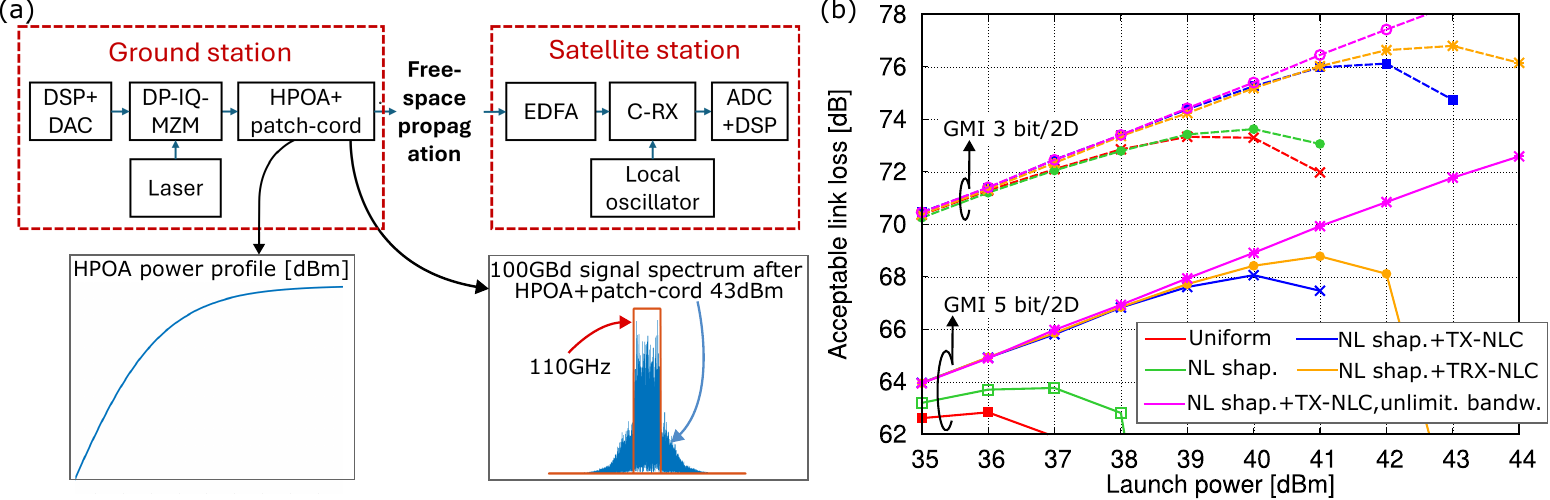}

\caption{\label{fig:fig1}(a) System setup, (b) system performance}

\vspace*{-2ex}
\end{figure}

\section{System Setup and Performance}

\vspace*{-1ex}

We evaluate the performance of the proposed techniques through numerical
simulations of the system shown in Fig.\,\ref{fig:fig1}(a). The
TX, located at the ground station, performs DSP and uses a DAC to
map information bits on symbols from dual polarization (DP) $M$-quadrature
amplitude modulation (QAM) constellation with uniform or probabilistically
shaped distribution. The symbols modulate a frequency-domain root-raised
cosine signal with roll-off $0.05$ and symbol rate $R=100$GBd. We
assume an electrical bandwidth of $110$GHz. The generated analog
signal modulates (through a dual-polarization in-phase and quadrature
Mach-Zehnder modulator DP-IQ-MZM) a continuous wave C-band laser.
The optical signal is amplified by a HPOA and launched into free space
with power $P$. In this paper, we model the HPOA as an EDFA of length
$30$m with attenuation $0.2\text{dB}/\text{km},$nonlinear coefficient
$3.6\text{W}^{-1}\text{km}^{-1}$, and dispersion $D=17\text{ps}/\text{nm}/\text{km}$,
with additional $3$m of SMF fibers for pigtails and patch cord. Fiber
propagation is simulated using the SSFM with a sufficiently large
number of steps. The power profile in the HPOA is assumed as the one
illustrated in the inset in Fig.\,\ref{fig:fig1}(a). After the
HPOA, the signal is launched into free space and received at the satellite
station. In this work, we model free-space propagation as an attenuation
of the signal by the loss $L$. At the receiver (RX) a low noise EDFA
(with noise figure $N_{F}=4$dB) amplifies the signal while also adding
ASE noise. The signal is received through a coherent RX (C-RX), and
is then processed by the ADC+DSP block, which also has a bandwidth
of $110$GHz. Assuming that both the EDFA gain and the link loss are
large, the overall signal-to-noise (SNR) at the RX is $P/\left(RLh\nu N_{F}\right)$,
where $h\nu$ is the photon energy. We assume that the TX laser and
LO linewidths are negligible, as these can be recovered with a CPR.

Figure \ref{fig:fig1}(b) shows the acceptable link loss as a function
of the launch power when the generalized mutual information (GMI)
is equal to either $3$bit/2D (solid lines) or $5$bit/2D (dashed
lines)---the maximum link loss which allows to achieve the desired
GMI, achievable with an ideal forward error correction code. Shaping
is obtained with rate $4.5$bit/2D on 64QAM or with rate $6.5$bit/2D
on 256QAM when the target GMI is $3$bit/2D or $5$bit/2D, respectively.
First, the figure compares the performance obtained with uniform with
shaped QAM using sphere shaping with optimized block length $N=4$
(green). The latter, labeled as NL shap., can be implemented with
a LUT of $256$ bits or $6144$ bits at the TX and $1280$ bits or
$36864$ bits at the RX, for the two shaping cases with $4.5$bit/2D
on 64QAM or $6.5$bit/2D on 256QAM, respectively. Overall, implementing
shaping with a LUT with negligible complexity allows to tolerate higher
link loss, up to 1dB for the case with target GMI $5$bit/2D. Nevertheless,
we believe that this nonlinear shaping gain might be improved by properly
tailoring shaping for the HPOA scenario. Next, the figure shows the
performance obtained adding NLPR on top of shaping with the LUT. The
ideal performance assuming very large device's bandwidth and NLPR
at the TX is shown as a reference (gray), and perfectly recovers nonlinear
effects (in the considered power region). However, when the electrical
bandwidth is limited to $110$GHz, the performance with NLPR at the
TX drops (magenta), still yielding an additional gain of up to $4$dB.
This gain further increases when the NLPR is splitted between TX and
RX ($\kappa=0.6$), at the expense of a modest increase in receiver
complexity. Overall, our results show that the link loss can increase
by $3.5$dB and $6$dB for the cases with target GMI $3$bit/2D or
$5$bit/2D using probabilistic shaping implemented with low-complexity
LUT combined with TX and RX NLPR. The RX-side NLPR can be omitted
with only a minor performance penalty.

\section{Conclusion}

\vspace*{-1ex}
This work proposed two complementary DSP techniques to mitigate the
nonlinearity arising in ground-stations HPOA for coherent ground-to-satellite
optical links. Properly shaping the signal at the ground station with
minimal complexity, the acceptable link loss increases by up to $5.2$dB
(GMI $5$bit/2D, shaping rate $6.5$ bit/2D 256QAM, $100$GBd signal,
with LUT-based shaping and TX-NLPR). The performance further improves
enabling up to $6$dB link loss with a small additional complexity
at the satellite station.

\section*{Acknowledgment}

\vspace*{-1ex}
This work was partially supported by the European Union under the
Italian National Recovery and Resilience Plan (NRRP) of NextGenerationEU,
partnership on \textquotedblleft Telecommunications of the Future\textquotedblright{}
(PE00000001 - program \textquotedblleft RESTART\textquotedblright ).


\begin{thebibliography}{10}
\small
\newcommand{\enquote}[1]{``#1''}

\bibitem{kaushal2016optical}
H.~Kaushal and G.~Kaddoum, \enquote{Optical communication in space: Challenges
  and mitigation techniques,} {\protect\JournalTitle{IEEE communications
  surveys \& tutorials}} \textbf{19}, 57--96 (2016).

\bibitem{li2022survey}
R.~Li, B.~Lin,et. al. \enquote{A survey on laser space
  network: terminals, links, and architectures,} {\protect\JournalTitle{IEEE
  Access}} \textbf{10}, 34815--34834 (2022).

\bibitem{boddeda2023achievableJLT}
R.~Boddeda, D.~R. Arrieta, et. al.
  \enquote{Achievable capacity of geostationary-ground optical links,}
  {\protect\JournalTitle{Journal of Lightwave Technology}} \textbf{41},
  3717--3725 (2023).

\bibitem{boddeda2024current}
R.~Boddeda, D.~R. Arrieta, et. al., \enquote{Current state,
  prospects, and opportunities for reaching beyond 100 gbps per carrier using
  coherent optics in satellite communications,} in \emph{ECOC 2024; 50th
  European Conference on Optical Communication,}  (VDE, 2024), pp. 406--409.

\bibitem{kotov2022high}
L.~V. Kotov, V.~Temyanko, et. al.,
  \enquote{High-energy single-frequency core-pumped er-doped fiber amplifiers,}
  {\protect\JournalTitle{Journal of Lightwave Technology}} \textbf{41},
  1526--1532 (2022).

\bibitem{kotov2014yb}
L.~V. Kotov, M.~E. Likhachev, et. al.,
  \enquote{{Y}b-free {E}r-doped all-fiber amplifier cladding-pumped at 976 nm
  with output power in excess of 100 {W},} in \emph{Fiber Lasers XI:
  Technology, Systems, and Applications,} , vol. 8961 (SPIE, 2014), pp.
  149--154.

\bibitem{matniyaz2020302}
T.~Matniyaz, F.~Kong, et. al., \enquote{302 {W}
  single-mode power from an {E}r/{Y}b fiber {MOPA},}
  {\protect\JournalTitle{Optics Letters}} \textbf{45}, 2910--2913 (2020).

\bibitem{nicholson2025high}
J.~W. Nicholson, A.~Grimes, et. al., \enquote{High-power fiber lasers for free-space optical
  communication uplinks,} {\protect\JournalTitle{IEEE Journal of Selected
  Topics in Quantum Electronics}}  (2025).

\bibitem{billault2023optical}
V.~Billault, S.~Leveque, et. al., \enquote{Optical coherent combining
  of high-power optical amplifiers for free-space optical communications,}
  {\protect\JournalTitle{Optics Letters}} \textbf{48}, 3649--3652 (2023).

\bibitem{ciaramella2024nonlinear}
E.~Ciaramella and G.~Cossu, \enquote{Nonlinear fiber effects in ultra-high
  power 10$\times$ 10 gbit/s {WDM-IM} free-space systems for satellite links,}
  {\protect\JournalTitle{Optics Express}} \textbf{32}, 7959--7968 (2024).

\bibitem{guiomar2022coherent}
F.~P. Guiomar, M.~A. Fernandes,  et. al., \enquote{Coherent free-space optical communications: Opportunities
  and challenges,} {\protect\JournalTitle{Journal of Lightwave Technology}}
  \textbf{40}, 3173--3186 (2022).

\bibitem{bocherer2015bandwidth}
G.~B{\"o}cherer, F.~Steiner, and P.~Schulte, \enquote{Bandwidth efficient and
  rate-matched low-density parity-check coded modulation,}
  {\protect\JournalTitle{IEEE Transactions on Communications}} \textbf{63},
  4651--4665 (2015).

\bibitem{civelli2022JLTBPS}
S.~Civelli, E.~Forestieri, and M.~Secondini, \enquote{On the nonlinear shaping
  gain with probabilistic shaping and carrier phase recovery,}
  {\protect\JournalTitle{Journal of Lightwave Technology}} \textbf{10},
  3046--3056 (2022).

\end{thebibliography}
\end{document}